\pdfoutput=1
\documentclass{article}
\usepackage{spconf,amsmath,graphicx,hyperref}

\usepackage{booktabs}
\usepackage{tikz}
\usepackage{pgfplots}
\pgfplotsset{width=8cm,compat=1.12}
\usepgfplotslibrary{fillbetween}

\usepackage{caption}
\usepackage{subcaption}

\renewcommand{\paragraph}[1]{\noindent\textbf{#1}\ }

\title{An investigation into the adaptability of a diffusion-based TTS model}
\name{Haolin Chen, Philip N. Garner}
\address{Idiap Research Institute, Martigny, Switzerland}
\begin{document}
%
\maketitle
\begin{abstract}

Given the recent success of diffusion in producing natural-sounding synthetic speech, we investigate how diffusion can be used in speaker adaptive TTS.  Taking cues from more traditional adaptation approaches, we show that adaptation can be included in a diffusion pipeline using conditional layer normalization with a step embedding.  However, we show experimentally that, whilst the approach has merit, such adaptation alone cannot approach the performance of Transformer-based techniques. In a second experiment, we show that diffusion can be optimally combined with Transformer, with the latter taking the bulk of the adaptation load and the former contributing to improved naturalness.
\end{abstract}
\begin{keywords}
Text-to-speech, speaker adaptation, diffusion model, conditional layer normalization
\end{keywords}


\section{Introduction}
\label{sec:intro}

Recent years have seen successful applications of adaptive text-to-speech (TTS) \cite{gst,adaspeech,metastylespeech,yourtts} to synthesize personalized voices for target speakers. In the typical scenario of adaptive TTS, a source acoustic model, which is usually trained on a large multi-speaker corpus, is adapted with few adaptation data to synthesize the desired voice. Concurrently, in the general field of acoustic modeling, deep generative models (DGMs) \cite{glowtts,vits,difftts} have demonstrated their superiority over other solutions in high-quality and fast synthesis. In particular, the more recent diffusion models \cite{difftts,diffsinger,priorgrad} have dominated this field in terms of intelligibility and naturalness.

Current research in adaptive TTS focuses on 1) enhancing the generalizability of the source model to various acoustic conditions and styles; as well as 2) improving the data and parameter efficiency of adaptation.
The first can further be categorized into 1) employing pluggable reference encoders to generate representations of the acoustic information and style on different semantic levels \cite{gst,adaspeech,generspeech}; 
and 2) ad-hoc designs of model structure that control desired features \cite{adaspeech,metastylespeech}. 
Furthermore, such adaptation techniques should be based on architectures with high synthesis quality, in which aspect diffusion-based acoustic models have surpassed their flow-based predecessors \cite{glowtts,flowtron}, while enjoying more flexibility in network design. Since diffusion models were first applied in TTS, many works \cite{priorgrad,diffgantts,prodiff} have demonstrated how to accelerate the generative process substantially to a speed similar to that of their fastest counterpart without much degradation of synthesis quality. 

In general, we are interested in parameter-efficient adaptation techniques for diffusion-based acoustic models that enhance their generalizability. Despite diffusion models having been well explored for generic acoustic modeling, few works have exploited them in adaptive TTS systems. Guided-TTS 2 \cite{guidedtts2}, the only diffusion-based adaptive TTS system we are aware of, utilizes diffusion with classifier guidance to adapt to diverse voices. However, the method lacks parameter efficiency for each target as all parameters of the diffusion model are finetuned during adaptation, and is not within the typical encoder-decoder framework.
Since parameter-efficient adaptation techniques exist for other architectures such as Transformer \cite{adaspeech}, and given the superior synthesis quality of diffusion models, such a method for diffusion, would be of great interest to the community, enabling both parameter-efficient and high-quality adaptation.

Based on the analyses above, we investigate the adaptability of diffusion-based acoustic models, with a special focus on parameter-efficient solutions. Specifically, our experiment is based on a typical diffusion denoiser network architecture, being a bidirectional dilated convolutional neural network. Inspired by observations from HMM-based adaptation, we propose introducing conditional layer normalization (CLN) to the denoiser network. Preliminary experimental results suggest that although it is viable to adapt the diffusion decoder, simply relying on adapting diffusion is not sufficient for high-quality adaptation; it also indicates inferior generalizability and adaptability of the denoiser. We further introduce adaptive Transformer layers as part of the decoder and observe the impact of adding different numbers of such layers to the adaptation quality. Our result shows that, while CLN in the denoiser network contributes to better speech quality and speaker similarity, it must be used in combination with adaptive Transformer layers to achieve usable adaptation quality. We conclude that for this particular type of diffusion model, its best use case in an adaptive TTS system is as a post-processing net that helps refine the detail of mel-spectrograms generated by a Transformer decoder.


\section{Adaptive diffusion decoder}
\label{sec:adaptive}

\subsection{Diffusion-based acoustic model}

In principle, diffusion models generate samples by denoising a sample from the prior distribution into real data through a diffusion process. Although taking different approaches, the learning problem of diffusion models can be expressed in terms of learning a denoiser network that predicts the noise in each diffusion step. The prevalent architecture of diffusion acoustic models comprises a Transformer-based phoneme encoder and a diffusion denoiser decoder. Here we mainly focus on the network design of the denoiser.

The most widely used structure of the denoiser network is the bidirectional dilated convolutional network \cite{difftts,diffsinger,priorgrad,prodiff}; other choices include the U-Net \cite{guidedtts2,gradtts}.
As depicted in Fig. \ref{fig:denoiser}, the denoiser takes the sample from the previous step as input to predict the noise in the reverse diffusion process conditioned on the encoded phoneme sequence $C_{text}$ and the step embedding $t$. 
The network mainly consists of an input convolution layer and $N$ convolutional blocks with residual connections and skip outputs, after which the skip outputs are accumulated to generate the final prediction through output convolution layers.

\subsection{Conditional layer normalization for denoiser}

Previous works \cite{adaspeech,metastylespeech} find that the layer normalization in the Transformer can greatly affect the output with light-weight adaptable scale vector $\gamma$ and bias vector $\beta$: $LN(x) = \frac{x-\mu}{\sigma} * \gamma + \beta$, where $\mu$ and $\sigma$ are the mean and variance of the input vector $x$, respectively. Furthermore, the two vectors can be generated by a small neural network conditioned on the speaker embedding, which can be finetuned when adapting to a new voice, and significantly reduce the number of parameters to be adapted while maintaining adaptation quality. Following \cite{adaspeech}, we refer to this module as conditional layer normalization (CLN).

\begin{figure}
\centering
\begin{subfigure}{.6\columnwidth}
  \centering
  \includegraphics[width=0.90\columnwidth]{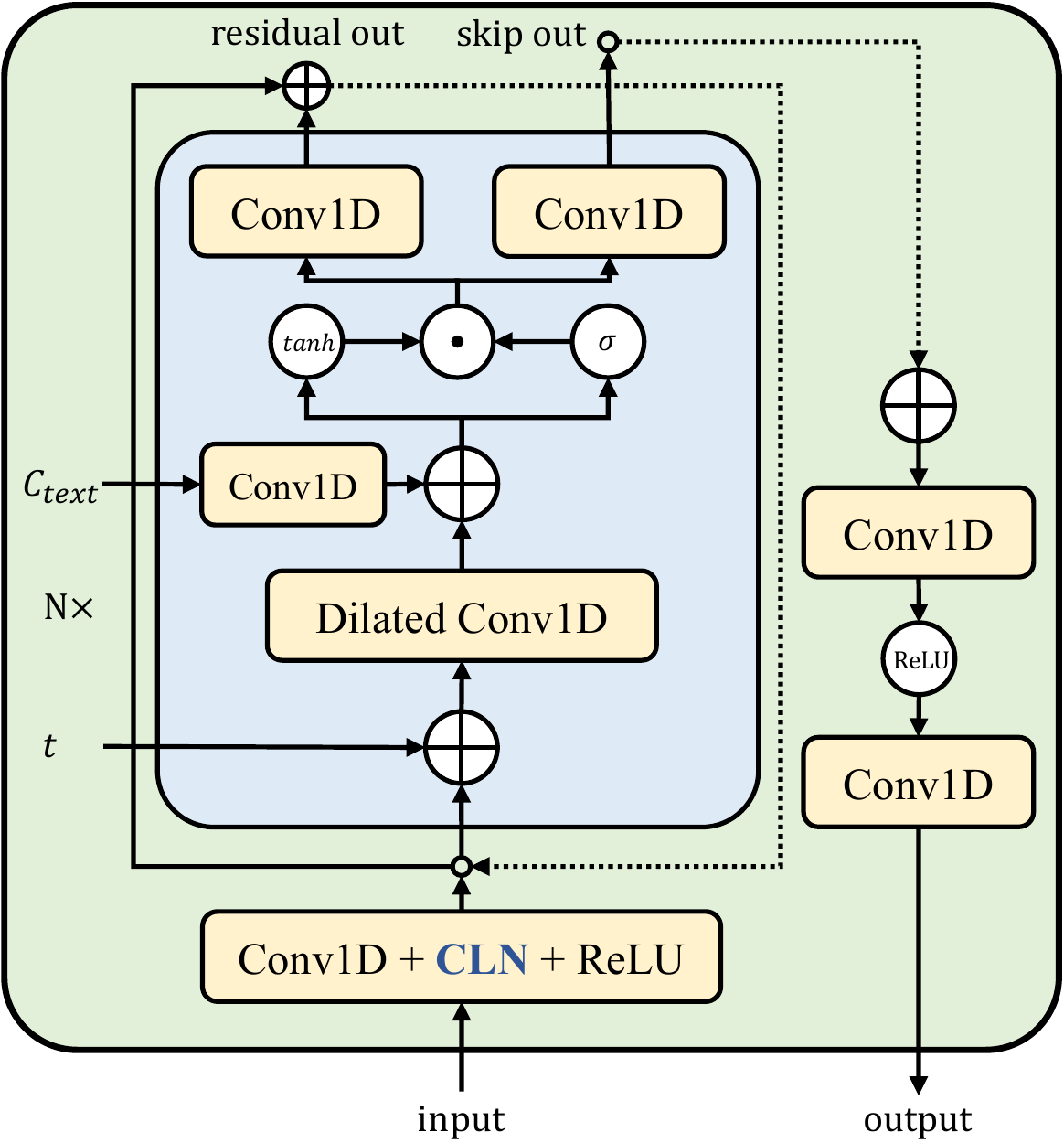}
  \caption{Denoiser network}
  \label{fig:denoiser}
\end{subfigure}%
\begin{subfigure}{.4\columnwidth}
  \centering
  \includegraphics[width=0.90\columnwidth]{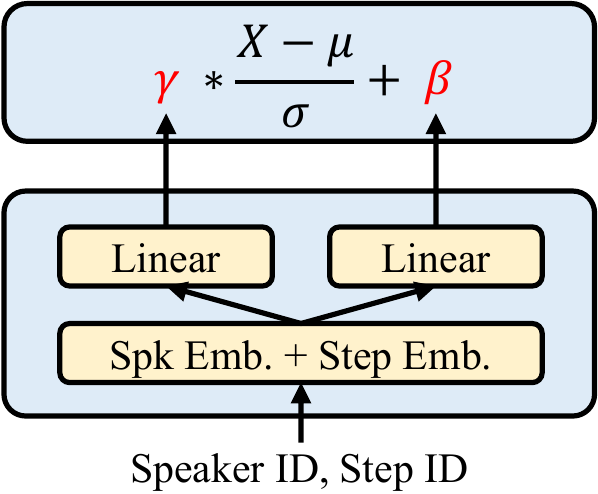}
  \caption{CLN}
  \label{fig:cln}
\end{subfigure}
\caption{Illustrations of the denoiser network and the conditional layer normalization (CLN).}
\label{fig:illustration}
\end{figure}

In particular, we are interested in integrating the CLN into the denoiser. Considering the application of the CLN in the Transformer, the operations take place on the whole hidden representation, and gradually change the prediction along the Transformer blocks.
Back to the denoiser network, the final prediction is collected by accumulating skip outputs from all convolution blocks, which is inspired by the idea of incorporating features at multiple levels to generate fine detail. Adding the CLN in or between these blocks makes it only apply to part of the features, which hinders the model from learning such hierarchical information. Moreover, the number of convolution blocks, $N$, is usually large ($\geq 12$), which hampers the parameter efficiency if the CLN is placed in every block.
After initial ad-hoc experiments that verified the hypotheses above, we place the CLN right after the 1-D convolution layer at the input, as in Fig. \ref{fig:denoiser}. The positioning is also inspired by HMM-based adaptation methods such as \cite{Leggetter1994}, in which the normalization takes place in the frequency (or cepstral) domain on the whole feature. 

The property of diffusion models that a single denoiser is used in all diffusion steps by being conditioned on the step embedding also makes it a parameter-efficient solution. Instead of being solely conditioned on the speaker embedding, the CLN takes the concatenation of both speaker and step embedding to control the strength of the operation, as is depicted in Fig. \ref{fig:cln}. Since the module is shared cross all diffusion steps, and the normalization should come into effect only when the sample is more refined rather than at the beginning, this mechanism enables the denoiser to automatically learn when to start functioning and at what strength during the whole reverse diffusion process. This is also inspired by \cite{DBLP:conf/cvpr/GuCBWZCYG22/diffusiontransformer}, in which a Transformer-based denoiser is conditioned on a step embedding through CLN. 

The following sections describe experiments to verify the system described thus far.






\subsection{Experiment settings}

\paragraph{Implementation details.} The model architecture of the diffusion decoder used in the experiments is based on PriorGrad \cite{priorgrad}. As an improved version of Diff-TTS \cite{difftts}, the first diffusion-based acoustic model, PriorGrad leverages a data-dependent prior, of which the mean and variance are phoneme-level statistics extracted from the dataset. Compared to ones using a standard Gaussian prior, PriorGrad offers better synthesis quality, higher parameter efficiency, and faster convergence. 
For the architecture behind the diffusion decoder, we implemented that of AdaSpeech \cite{adaspeech}, including the phoneme encoder, the acoustic condition modeling module, and the variance adapter. 
Our implementation is based on the open-source software \footnote{NATSpeech: https://github.com/NATSpeech/NATSpeech} \footnote{NeuralSpeech: https://github.com/microsoft/NeuralSpeech} of the two models. 
We use the diffusion decoder with 12 convolution blocks, 128 residual channels and 3.5M parameters proposed in \cite{priorgrad}. Other model configurations follow corresponding parts of AdaSpeech and PriorGrad unless otherwise stated. 
The total number parameters to be finetuned for the diffusion decoder is 0.131M, compared to 1.184M for AdaSpeech.

\paragraph{Data.} We train the source model on two clean subsets train-clean-100 and train-clean-360 of LibriTTS dataset \cite{libritts}, a multi-speaker TTS corpus, totaling 1151 speakers and 245 hours of speech. For evaluation, we select 11 speakers (7 females and 4 males) with different accents from VCTK \cite{vctk} 
following the practice in \cite{yourtts}. For each speaker, 10 utterances with the same transcripts across all speakers are randomly selected as the test set. The preprocessing of speech and text data follows AdaSpeech except using a sampling rate of 22,050 Hz.

\paragraph{Training, adaptation and inference.} Following AdaSpeech, the training process comprises two stages in which the numbers of steps are 200K and 100K, respectively. The models are trained on one NVIDIA RTX 3090 GPU using a batch size of 50,000 speech frames.
For the diffusion decoder, a beta schedule with 400 steps is used for both training and inference. We use the speaker-independent prior calculated on the whole training set. Other hyperparameters follow \cite{priorgrad} unless otherwise stated. During adaptation, the model is finetuned using 10 utterances of the target speaker for 2000 steps using a fixed learning rate of $2\times10^{-4}$, while only the speaker embedding and the CLN are optimized. In the inference process, a HiFi-GAN vocoder \cite{hifigan} trained on VCTK is used to synthesize waveforms from the generated spectrogram.

\subsection{Objective evaluation}

For preliminary evaluation, we employ MOSNet \cite{mosnet}, a neural network-based objective assessment tool for speech quality that generates machine rated MOS (mean opinion score), and a pretrained speaker verification model provided by SpeechBrain \cite{speechbrain} which calculates the cosine similarity (CS) between speaker embeddings of the generated sample and the reference. The cosine score ranges from 0 to 1; higher score means higher speaker similarity to the reference. 
We found the scores from the two automatic assessment tools were consistent with our subjective judgment, and will conduct human evaluation for the final settings.

We compare the performance among the following settings: 1) GT mel + Vocoder, using the ground truth mel-spectrograms to synthesize waveforms with the HiFi-GAN vocoder; 2) AdaSpeech, the Transformer-based adaptive acoustic model with CLN applied to the decoder; 3) Enc + DiffDec (decoder), a baseline system without the CLN in the denoiser which  finetunes the whole decoder during adaptation as an upper bound; 4) Enc + DiffDec (spk emb), with the same architecture as the previous one but only finetuning the speaker embedding as a lower bound; 5) Enc + AdaDiffDec, our proposed system with the CLN in the denoiser that is finetuned with the speaker embedding during adaptation.

\begin{table}[tbp]
  \centering
  \caption{The MOSNet and cosine similarity (CS) scores.}
    \begin{tabular}{c|l|c|c}
    \toprule
    \textbf{\#} & \textbf{Model} & \textbf{MOSNet\,($\uparrow$)} & \textbf{CS\,($\uparrow$)} \\
    \midrule
    1     & \textit{GT mel + Vocoder} & 4.10  & 0.96  \\
    \midrule
    2     & \textit{AdaSpeech} & 3.78  & 0.52  \\
    \midrule
    3     & \textit{Enc + DiffDec (decoder)} & 3.80  & 0.50  \\
    4     & \textit{Enc + DiffDec (spk emb)} & 3.42  & 0.20  \\
    \midrule
    5     & \textit{Enc + AdaDiffDec} & 3.58  & 0.22  \\
    \bottomrule
    \end{tabular}%
  \label{tab:obj}%
\end{table}%

\subsection{Results and analyses}

The MOSNet and cosine similarity results are shown in Table \ref{tab:obj}. It can be observed that: 1) adapting the whole diffusion decoder (\#3) results in the best speech quality and speaker similarity among all settings, achieving similar performance to AdaSpeech (\#2); 2) only finetuning the speaker embedding (\#4) results in poor performance; 3) our proposed method (\#5) only slightly outperforms baseline (\#4), nevertheless it is much worse than finetuning the whole decoder (\#3) and AdaSpeech (\#2).

The results indicate that simply relying on adapting the CLN in the denoiser is not sufficient for achieving a reasonable adaptation quality. Furthermore, our test listening suggests that some of the samples synthesized by three diffusion-based systems (\#3-5) are not intelligible, which explains their inferiority to AdaSpeech demonstrated by objective tests. It also implies that the diffusion decoder is sensitive to out-of-domain input, therefore has poor generalizability and adaptability. This is very interesting since it is capable of synthesizing very high-quality and natural speech as a generic acoustic model. The phenomenon suggests that further efforts should be made to improve the adaptability of the system.


\section{Adapting diffusion: a good choice?}
\label{sec:choice}

Our preliminary experimental result demonstrated that the previously proposed system does not achieve usable adaptation quality, which suggests that solely adapting the diffusion decoder may not be a good choice; other components need to be introduced to the system to improve the adaptation performance while taking advantage of the high-quality synthesis of the diffusion decoder. 





Given the fact that Transformer-based adaptive TTS systems have achieved decent adaptation quality, we consider adding Transformer layers with CLN before the diffusion decoder to construct a decoder with mixed architecture. 
The method is inspired by a common practice of utilizing DGMs as post-processing nets (post-net) in acoustic models \cite{vits,portaspeech} to refine the over-smoothed sample generated by VAE (variational auto-encoder) or Transformer. Much improved adaptation quality is expected provided that the diffusion decoder works as a post-net that refines the output of AdaSpeech. We are especially interested in how much performance the additional Transformer layers can bring and the difference between adapting both the Transformer decoder and the diffusion decoder and adapting the Transformer decoder alone.

\subsection{Experiments and evaluation}

The model configurations in this setting are a grid search combining the following two factors: 1) the number of additional Transformer decoder layers from 0 to 4, where 4 corresponds to the full Transformer decoder; and 2) whether or not to use the CLN in the diffusion denoiser. Since there are more than 10 systems to compare, we first conduct the objective evaluation as previous experiments.  We then further conduct subjective listening tests to verify the findings from objective test results. We expect that more Transformer layers result in better speech quality and speaker similarity. However, it is more important to observe the impact of the CLN in diffusion on the performance in such settings.

For subjective listening tests, 10 raters were involved to rate the MOS for naturalness and SMOS for similarity of 10 samples for each system. The test utterances were randomly selected from those used in objective tests, covering most speakers or accents. Note that the test utterances are the same across all systems. We select the two settings with 4 Transformer decoder layers, which are equivalent to using the diffusion decoder as a post-net on top of the AdaSpeech and are expected to have the best performance, and compare them with the vocoder-synthesized ground truth and AdaSpeech.

\subsection{Results}

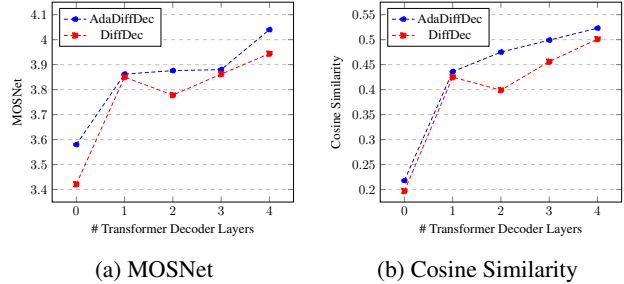
\begin{figure}[tbp]
\centering
\begin{subfigure}{.5\columnwidth}
  \centering
  \begin{tikzpicture}[scale=0.5]
\begin{axis}[
    xlabel={\# Transformer Decoder Layers},
    ylabel={MOSNet},
    xmin=-0.5, xmax=4.5,
    ymin=3.35, ymax=4.15,
    xtick={0,1,2,3,4},
    ytick={3.4, 3.5, 3.6, 3.7, 3.8, 3.9, 4.0, 4.1},
    legend pos=north west,
    ymajorgrids=true,
    grid style=dashed,
]

\addplot[
    color=blue,
    mark=*,
    densely dashed
    ]
    coordinates {
    (0,3.580)
    (1,3.862)
    (2,3.876)
    (3,3.880)
    (4,4.040)
    };
    
\addplot[
    color=red,
    mark=square*,
    densely dashed
    ]
    coordinates {
    (0,3.421)
    (1,3.850)
    (2,3.778)
    (3,3.862)
    (4,3.944)
    };
    \legend{AdaDiffDec,DiffDec}
    
\end{axis}
\end{tikzpicture}
  \caption{MOSNet}
  \label{fig:obj-mosnet}
\end{subfigure}%
\begin{subfigure}{.5\columnwidth}
  \centering
  \begin{tikzpicture}[scale=0.5]
\begin{axis}[
    xlabel={\# Transformer Decoder Layers},
    ylabel={Cosine Similarity},
    xmin=-0.5, xmax=4.5,
    ymin=0.175, ymax=0.575,
    xtick={0,1,2,3,4},
    ytick={0.20, 0.25, 0.30, 0.35, 0.40, 0.45, 0.50, 0.55},
    legend pos=north west,
    ymajorgrids=true,
    grid style=dashed,
]

\addplot[
    color=blue,
    mark=*,
    densely dashed
    ]
    coordinates {
    (0,0.218 )
    (1,0.436 )
    (2,0.475 )
    (3,0.499 )
    (4,0.523 )
    };
    
\addplot[
    color=red,
    mark=square*,
    densely dashed
    ]
    coordinates {
    (0,0.197 )
    (1,0.425 )
    (2,0.399 )
    (3,0.456 )
    (4,0.501 )
    };
    \legend{AdaDiffDec,DiffDec}
    
\end{axis}
\end{tikzpicture}
  \caption{Cosine Similarity}
  \label{fig:obj-secs}
\end{subfigure}
\caption{The MOSNet and cosine similarity scores of settings with different number of Transformer decoder layers.}
\label{fig:obj}
\end{figure}

The results of the objective evaluation are shown in Figure \ref{fig:obj}, where the results of the two metrics are displayed in separate plots. Several observations include: 1) the additional Transformer layers significantly improve the performance compared to only using the diffusion decoder; 2) in general, both speech quality and speaker similarity are improved with increasing number of Transformer layers; 3) adding CLN to the denoiser results in better performance in terms of both metrics in all settings, however, the difference of speaker similarity narrows when the number of Transformer layers is high. The subjective test results are shown in Table \ref{tab:subj}, where ``AS'' stands for AdaSpeech. It can be seen that 1) the diffusion decoder on top of the Transformer decoder (\#3,4) significantly improves both perceptual speech quality and speaker similarity compared to AdaSpeech (\#2) which only uses the Transformer decoder; 2) the CLN in the diffusion decoder further improves the two scores, making \#3 the best among all systems; 3) the improvement of speech quality the CLN brings is more than that of speaker similarity, which only shows a slight difference. 


Overall, it has been demonstrated that, despite the CLN in the denoiser network contributing to higher adaptation quality, it must be used with adaptive Transformer layers to achieve usable performance.  The adaptability of the model mainly relies on the adaptive Transformer layers, which suggests the inferior generalizability and adaptability of the diffusion denoiser compared to the Transformer.

\begin{table}[tbp]
  \centering
  \caption{The MOS and SMOS scores with 95\% confidence.}
    \begin{tabular}{c|l|c|c}
    \toprule
    \textbf{\#}    & \textbf{Model} & \textbf{MOS ($\uparrow$)} & \textbf{SMOS ($\uparrow$)} \\
    \midrule
    1     & \textit{GT mel + Vocoder} & 4.77 ± 0.10 & 4.99 ± 0.02 \\
    \midrule
    2     & \textit{AdaSpeech} & 2.38 ± 0.19 & 2.94 ± 0.20 \\
    \midrule
    3     & \textit{AS + AdaDiffDec} & 3.02 ± 0.19 & 3.24 ± 0.20 \\
    4     & \textit{AS + DiffDec} & 2.84 ± 0.19 & 3.16 ± 0.21 \\
    \bottomrule
    \end{tabular}%
  \label{tab:subj}%
\end{table}%


\label{sec:discussion}


\section{Conclusion}
\label{sec:conclusion}

In this paper, we conducted an investigation into the adaptability of a typical diffusion-based acoustic model under parameter-efficient settings. We proposed the conditional layer normalization for the denoiser network and tested its effectiveness for speaker adaptation. We demonstrated that, while it is feasible to adapt the diffusion decoder by this method, it must be used in combination with adaptive Transformer layers to achieve usable adaptation quality. This suggests that the diffusion is less generalizable and adaptable than a Transformer. Future works in this field should focus on improving the diffusion model in the above aspects, or utilize the diffusion model as a post-net that refines the mel-spectrograms generated by other adaptable components.



\section{Acknowledgments}
\label{sec:ack}

This project received funding under NAST: Neural Architectures for Speech Technology, Swiss National Science Foundation grant \href{https://data.snf.ch/grants/grant/185010}{185010}.

\vfill\pagebreak

\bibliographystyle{IEEEbib}
\bibliography{chl-ref-short}

\end{document}